# Experimental demonstration of birefrigent transformation optics devices

Vera N. Smolyaninova [1], H. Kurt Ermer [1], Alex Piazza [1], David Schaefer [1], and

Igor I. Smolyaninov [2]

[1] *Department of Physics Astronomy and Geosciences, Towson University,*

*8000 York Rd., Towson, MD 21252, USA*

[2] *Department of Electrical and Computer Engineering, University of Maryland,*

*College Park, MD 20742, USA*

**Transformation optics (TO) has recently become a useful methodology in the design of unusual optical devices, such as novel metamaterial lenses and invisibility cloaks. Very recently Danner *et al.* [1] have suggested theoretical extension of this approach to birefrigent TO devices, which perform useful and *different* functions for mutually orthogonal polarization states of light. Theoretical designs which operate as invisibility cloak for one polarization while behaving as a Luneburg lens for another orthogonal polarization have been suggested. Here we report the first experimental realization of similar birefrigent TO designs based on lithographically defined metal/dielectric waveguides. Adiabatic variations of the waveguide shape enable control of the effective refractive indices experienced by the TE and TM modes propagating inside the waveguides. We have studied wavelength and polarization dependent performance of the resulting birefrigent TO devices. These novel optical devices considerably extend our ability to control light on submicrometer scales.**

Recent progress in metamaterial and transformation optics (TO) research gave rise to such fascinating devices as perfect lenses [2], invisibility cloaks [3-5], perfect absorbers [6], and numerous other unusual electromagnetic devices. Very recently Danner *et al.* [1] have suggested theoretical extension of this approach to birefrigent TO devices, which perform useful and *different* functions for mutually orthogonal polarization states of light. This development has changed common notion that birefringence constitutes an undesirable obstacle in metamaterial and TO research. Instead, Danner *et al.* have demonstrated theoretically that birefringence offers an additional degree of freedom, which can benefit an optical designer. For example, they have suggested electromagnetic devices which operate as invisibility cloak for one polarization while behaving as a Luneburg lens for another orthogonal polarization of light. This and some other examples have been described in detail in Ref.[1]. Here we report on the first experimental realization of such birefrigent TO devices, which operate in the visible frequency range. Our designs are based on lithographically defined metal/dielectric waveguides. Adiabatic variations of the waveguide shape enable control of the effective refractive indices experienced by the TE and TM modes propagating inside the waveguides. We present an experimental realization of a device which operates as a Luneburg lens for TM polarized light, while behaving as spatial (directional) filter for TE polarized light. In the second design a Luneburg lens for TM light has a cloaking potential in its center for TE light. Our experimental designs appear to be broadband, which has been verified in the 480-633 nm range. These novel optical devices considerably extend our ability to control light on submicrometer scales.

Our approach is based on the recent demonstration that metamaterial parameter distribution required for cloaking and other TO-based designs can be emulated by adiabatic changes of shape of a 2D metal-dielectric-metal optical waveguide [7]. Cloaking performance of the waveguide geometry considered in [7] exhibits almost no polarization dependence, which has been verified experimentally. The fundamental mode of the metal-dielectric-metal waveguide considered in [7] is a plasmon mode, which has no cutoff, but has extremely short propagation length at $\lambda$=500 nm. Therefore, existence of this mode does not affect experimental results obtained in [7]: the other TE and TM polarized waveguide modes experience very similar effective refractive index distribution inside the tapered waveguide. On the other hand, adiabatically changing dielectric waveguide which does not have a top metallic layer behaves very differently with regards to polarization of illuminating light, since its fundamental TM mode has a long propagation length.

Let us consider a three-layer waveguide geometry which is shown schematically in the inset in Fig.1. Assuming adiabatic changes of the waveguide thickness, the wavevector $k$ of the guided mode can be calculated as a function of light frequency $\omega$ and waveguide thickness $d$ for TE and TM polarized modes, resulting in the definition of effective refractive index $n_{eff}=k\omega/c$ for both polarizations. The implicit equations defining $k$ as a function of $\omega$ can be calculated via boundary conditions at two interfaces as follows:

$$\left(\frac{k_1}{\varepsilon_m}-\frac{ik_2}{\varepsilon}\right)\left(k_3-\frac{ik_2}{\varepsilon}\right)e^{-ik_2d}=\left(\frac{k_1}{\varepsilon_m}+\frac{ik_2}{\varepsilon}\right)\left(k_3+\frac{ik_2}{\varepsilon}\right)e^{ik_2d} \qquad (1)$$

for the TM, and

$$\left(k_1-ik_2\right)\left(k_3-ik_2\right)e^{-ik_2d}=\left(k_1+ik_2\right)\left(k_3+ik_2\right)e^{ik_2d} \qquad (2)$$

for the TE polarized guided modes, where the vertical components of the wavevector $k_i$ are defined as:

$$k_1 = \left( k^2 - \varepsilon_m \frac{\omega^2}{c^2} \right)^{1/2}, \; k_2 = \left( \frac{\omega^2}{c^2} \varepsilon - k^2 \right)^{1/2}, \text{ and } k_3 = \left( k^2 - \frac{\omega^2}{c^2} \right)^{1/2} \qquad (3)$$

in metal, dielectric, and air, respectively. In the limit $\varepsilon_m \to -\infty$ Eqs.(1,2) are simplified as follows:

$$k_3 \cos k_2 d = k_2 \sin k_2 d \qquad \text{for TM,} \quad \text{and}$$

$$k_3 \sin k_2 d = -k_2 \cos k_2 d \qquad \text{for TE} \qquad (4)$$

Solutions of eqs.(4) are plotted in Fig.1, which shows the resulting effective refractive indices for both polarizations. Effective birefringence for the lowest guided TM and TE modes appears to be very strong at waveguide thickness $d$<0.4 μm, and both polarizations demonstrate strong index dependence on the waveguide thickness. This behavior can be used in building non-trivial birefrigent TO devices if a waveguide thickness as a function of spatial coordinates $d(r)$ may be controlled with enough precision. For example, a modified Luneburg lens [8] with radial refractive index distribution

$$n = \sqrt{1 + f^2 - \left( r/a \right)^2} \, / f \qquad \text{for } r < a \qquad (5)$$

in which refractive index varies from $n(0)= \sqrt{1+f^2} \, / f$ to $n(a)$=1 is easy to realize for TM polarized light based on the theoretical data plotted in Fig.1. Theoretical performance of such a lens for $f$=1 is presented in Fig.2(a) based on COMSOL Multiphysics simulations. On the other hand, the same $d(r)$ profile produces a different refractive index distribution for TE polarized light, which changes from $n(0)$~1.41 to $n(a)$~0. Due to near zero effective refractive index near the device edge, a Luneburg lens for TM light will operate as a spatial (directional) filter for TE light, as shown in Fig.2(b). This result is natural since most of TE light must experience total reflection from the interface between air ($n$=1) and the lens edge ($n$~0) coming from the medium with higher refractive index.

We have developed a lithography technique which enables such precise *d(r)* shape control of the dielectric photoresist on gold film substrate. In traditional lithographic applications for the best results of pattern transfer the profile of resist should be rectangular or even with overhang. Therefore, several well known methods are employed to achieve the sharpest edge possible. Our purpose is different. We want to create a more gradual edge profile. This can be done by disregarding typical precautions employed to make the edges sharp. To produce gradual decrease of photoresist thickness (Shieply S1811 photoresist having refractive index n~1.5 was used for device fabrication) several methods have been used. Instead of contact printing (when mask is touching the substrate), we used soft contact mode (with the gap between the mask and the substrate). This allows for the gradient of exposure due to the diffraction at the edges, which leads to a gradual change of thickness of the developed photoresist. Underexposure and underdevelopment were also used to produce softer edges. It was also possible to produce donut shape patterns using slightly larger exposure and development time than for the circular patterns. Examples of so formed TO devices are presented in Figs.2(c) and 5(b). Variations of height of the photoresist patterns provide efficient control of the TM and TE refractive index within the device. As demonstrated by Fig.2(e), we were able to fabricate photoresist patterns which almost ideally fit the modified Luneburg lens profile described by eq.(5). However, we have observed that such a close fit was not really necessary for the best focusing performance. Both devices shown in Fig.2(c) performed equally well.

Experimental images in Figs.3,4 demonstrate measured performance of the designed birefrigent TO device. In these experiments a near-field scanning optical microscope (NSOM) fiber tip was brought in close proximity to the arrays of lithographically formed TO devices and used as an illumination source. Almost diffraction-limited ($\sim 0.7\lambda$) focusing of 515 nm light (Fig.3) emitted by the fiber tip (seen on the left) clearly demonstrates Luneburg lens-like focusing behavior of our birefrigent TO device arrays for TM polarized light. Note that previous attempts of TO lens fabrication [9,10] were only able to achieve focusing spot size of the order of $3\lambda$, so that an order of magnitude improvement in TO lens quality has been achieved in our experiments. Comparison of theoretical and experimental images performed in Fig. 3(d) demonstrates excellent agreement between theory and experiment for both polarizations (artificial color scheme used to represent experimental images in Fig.3(d) has been chosen to better highlight this close match). Theoretical images in Fig.3(d) were calculated by taking into account real device profile shown in Fig.2(d), so they differ slightly from Figs.2(a,b). We should also note that the effective refractive index for TE wave should be an imaginary number for d<80 nm (see Fig.1). However, experimentally measured device profile shown in Fig.2(d) indicates that the effective TE refractive index is imaginary only within an extremely thin rim at the very edge of our device. From Fig.2(d) the thickness of this rim can be estimated as no more than 20 nm, or 1/25 of light wavelength. While such a thin layer will behave as a tunneling barrier for TE light, transmission of such a thin layer is quite large, as is obvious from experimental and theoretical data presented in Fig.3(d).

Measured wavelength and polarization dependencies of light intensity in the focal spot of the lenses shown in Fig.4 further validate our design. Birefrigent TO device operation at 633 nm demonstrates broadband performance of our design: device performance at 633 nm is similar to focusing performance at 515 nm, which is presented in Fig.3(a). Polarization dependence in Fig.4(b) has been measured for several lenses (marked as #1, 2, and 3) located in the same horizontal row as the NSOM fiber tip used as a source. A succession of images similar to the one shown in Fig.4(a) were taken through a rotated linear polarizer (see Fig.4(c)). While the fiber tip emits unpolarized light, polarization response of the image produced by a distant TO device can be clearly separated into TM and TE contributions with respect to the plane of incidence of the source light. While differences in polarization response between individual lenses are present, and can be accounted for by device shape imperfections, overall behavior demonstrates excellent agreement with theory.

Another interesting option provided by strong birefringence of our device geometry at waveguide thicknesses $d<0.4$ μm, is a possibility to produce a waveguide device which would operate as a slightly perturbed Luneburg lens for TM polarized light, while exhibiting an approximate semi-classical cloaking Hamiltonian [11] for TE polarized light inside the device. As can be seen from Fig.1, variations of waveguide thickness in the 0.1-0.3 μm range lead to very strong variations of TE effective refractive index in the $0<n_{eff}<1.4$ range, while keeping the TM refractive index approximately constant at $n_{eff}>1.3$. Thus, as shown in Fig.5(a,b), a broadband cloaking geometry described in ref.[7] may be replicated near the device center for TE polarized light only, while keeping Luneburg lens-like refractive index distribution for the TM polarization. TE guided mode behavior in an uncoated tapered dielectric waveguide shown in Fig.5(a) is somewhat similar to the TE mode behavior inside an air-filled waveguide between two gold-coated surfaces used in ref.[7]. In both cases the tapered waveguides exhibit well-defined cutoffs so that the *d*-dependences of the effective *n* are similar. The dispersion law of TE mode inside gold-coated waveguide may be given by the following approximate expression:

$$\frac{\omega^2}{c^2} = k_r^2 + \frac{k_\phi^2}{r^2} + \frac{\pi^2 m^2}{d(r)^2} \ , \tag{6}$$

where *m* is the transverse mode number [7]. A photon launched into the *m*-th mode of the waveguide stays in this mode as long as *d* changes adiabatically. Thus, semi-classical 2D cloaking Hamiltonian (dispersion law) introduced in [11]:

$$\frac{\omega^2}{c^2} = k_r^2 + \frac{k_\phi^2}{(r-b)^2} = k_r^2 + \frac{k_\phi^2}{r^2} + k_\phi^2 \frac{b(2r-b)}{(r-b)^2 r^2} \tag{7}$$

may be emulated approximately using lithographically defined *d(r)* profile. Small values $n_{eff}<1$, which are necessary for cloaking behavior are obtained due to the waveguide cut-off observed as $d\rightarrow 0$ (see eq.(6)). 2D cloaking behavior has been indeed observed in such a waveguide [7], followed by its extension to 3D cylindrical geometry by Tretyakov *et al.* [12]. In our case similar cut-off behavior can be produced for the TE polarization only as demonstrated in Fig.1. Near the cut-off $k\rightarrow 0$ and $k_3$ becomes imaginary (see eq.(3)) similar to its behavior in the metal-dielectric-metal waveguide. Therefore, the TE dispersion law becomes similar to eq.(6) near the cut-off. It can be approximated as

$$\frac{\omega^2}{c^2} = k_r^2 + \frac{k_\phi^2}{r^2} + \frac{\pi^2 m^2}{n_d^2 d(r)^2} \ , \tag{8}$$

where $n_d$ is the refractive index of the photoresist. An estimate using eq.(8) differs from the numerically calculated cut-off (shown in Fig.1) by ~35%.

An array of the birefrigent TO devices described above has been fabricated and tested, as has been shown in Figs.5(b) and 6(c). Similar to polarization experiments in Fig.4, unpolarized light from the NSOM fiber tip may be separated into the TM and TE components for a distant TO device with respect to the plane of incidence of the source light. Thus, by viewing the optical field distribution through a linear polarizer oriented as shown in Fig.6(c), we can visualize the TE response of the devices marked as #1 and #2 in the image. This response demonstrates a good agreement with theoretical image shown in Fig.6(a). On the other hand, optical field distribution inside devices #3 and #4 viewed through the same linear polarizer has considerable TM component. In agreement with our theoretical design, these devices (which are identical to devices #1 and #2) do behave like Luneburg lenses for TM polarized light (compare with theoretical simulations shown in Fig.6(b)). Thus, similar to devices proposed in ref.[1], we have realized a TO device, which operates as a slightly perturbed Luneburg lens for TM polarized light, while exhibiting an approximate semi-classical cloaking Hamiltonian [11] for TE polarized light inside the device.

In conclusion, we have reported the first experimental realization of birefrigent TO devices, which perform different functions for mutually orthogonal polarization states of light. Using effective birefringence of a lithographically formed dielectric waveguide on a metal substrate, we have created a Luneburg lens for TM polarized light, which behaves as a spatial (directional) filter for TE polarized light. In the second design a Luneburg lens for TM light exhibits an approximate semi-classical cloaking Hamiltonian [11] for TE polarized light inside the device. Our technique opens up an additional degree of freedom in optical design and considerably improves our ability to manipulate light on submicrometer scale.

**Acknowledgements**

This research was supported by the NSF grants DMR-0348939 and DMR-1104676. We are grateful to J. Klupt and T. Adams for experimental help.

**Figure Captions**

**Figure 1.** Calculated birefringence of the effective refractive index as a function of thickness $d$ of the dielectric layer deposited onto the surface of ideal metal. The inset shows the dielectric waveguide geometry. The waveguide thickness $d$ is assumed to vary adiabatically.

**Figure 2**. (a) Theoretical simulations of a waveguide-based Luneburg lens for TM polarized light using COMSOL Multiphysics. In these simulations the lens diameter is set to 1. (b) The same device acts as a spatial (directional) filter for TE polarized light. (c) AFM images of various fabricated photoresist patterns which have been used to realize birefrigent TO devices presented in (a,b). The insets show 3D representations of their shapes. (d) Measured photoresist height variations near the edge of the left device shown in (c) along the gray line. This height variation provides necessary means to control the effective refractive index for TE and TM polarized light. (e) Measured photoresist height variations of the right device shown in (c) along the green line fitted to a modified Luneburg lens described by eq.(5). The fit is shown in red.

**Figure 3.** Focusing behavior of arrays of 6 μm diameter (a) and 2 μm diameter (b) birefrigent TO devices for TM polarized light. Almost diffraction-limited focusing of 515 nm light emitted by a tapered fiber tip (seen on the left) is clearly visible in these microscope images. The scale bar length is indicated in both images. Additional white light illumination was used to highlight device positions. (c) Experimentally measured cross section of the focal spot of the TO device. (d) Digital zoom of the measured field distributions inside the device for TM and TE polarized light is shown next to theoretical simulations, which take into account real device shape. Artificial coloring scheme is used to differentiate between the signal and illuminating light.

**Figure 4.** (a) Birefrigent TO device operation at 633 nm demonstrates broadband performance of our design: device performance at 633 nm is similar to focusing performance at 515 nm, which is presented in Fig.3(a). (b) Polarization dependence of light intensity in the focal spot for several devices marked as #1, #2 and #3 in Fig.4(a). (c) A succession of images similar to the one shown in Fig.4(a) taken through a rotating linear polarizer. The polarization angle is marked in the corner of each image. While the fiber tip emits unpolarized light, polarization response of the image produced by a distant TO device can be clearly separated into TM and TE contributions with respect to the plane of incidence of the source light.

**Figure 5.** (a) Profile variation $d(r)$ around the center of a donut-shaped dielectric waveguide fabricated on a gold film surface provide a cloaking potential similar to the one described in Ref.[7] for the TE polarized light only. (b) AFM image of a donut-shaped waveguide made of Shieply S1811 photoresist. The inset shows 3D representation of its shape.

**Figure 6**. COMSOL Multiphysics simulations of TE (a) and TM (b) light propagation inside the donut-shaped device show that TE light does not penetrate the "cloaked" area in the middle, while Luneburg lens-like performance for TM light is kept mostly intact. (c) Image of the array of donut-like TO devices obtained through a linear polarizer, which is oriented as shown by the arrow. While TE polarized light distribution in the central row (in front of the NSOM tip) matches theoretical simulations shown in (a), the TM component of light incident on other devices shows evidence of Luneburg lens-like focusing. (d,e,f) Enlarged images of a single device (d) showing field propagation around the device center for TE polarized illumination (e), and focusing for TM

polarized illumination (f), respectively. These images were obtained on a single device as a function of polarization angle rotation.

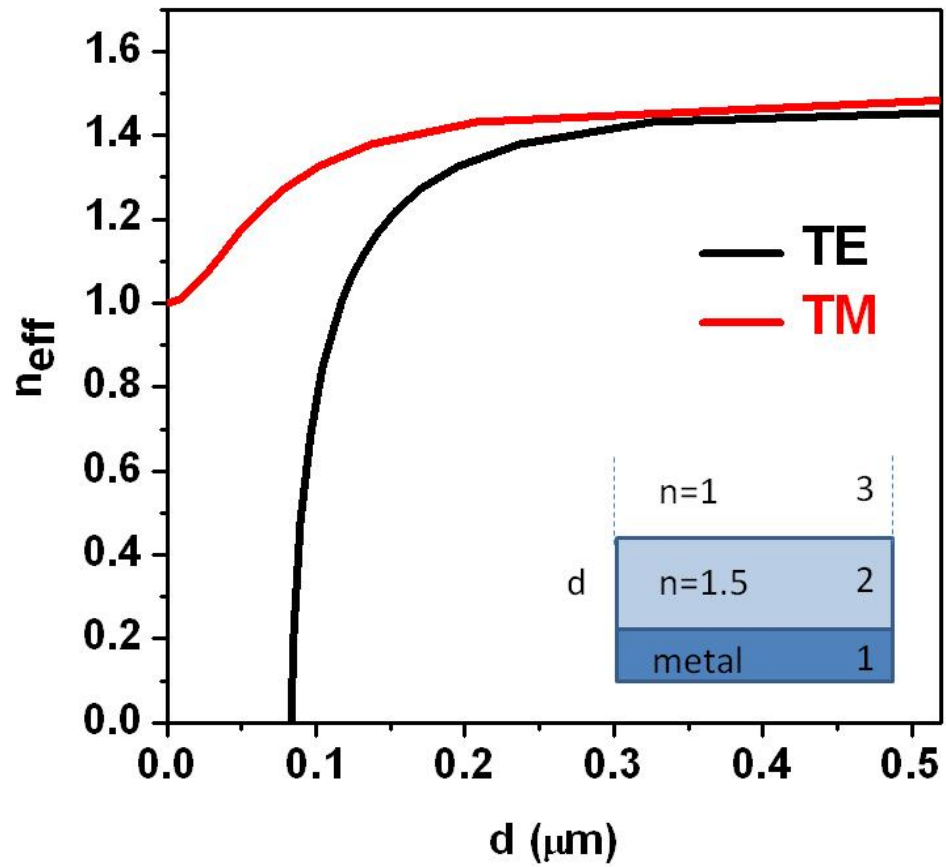

Fig.1

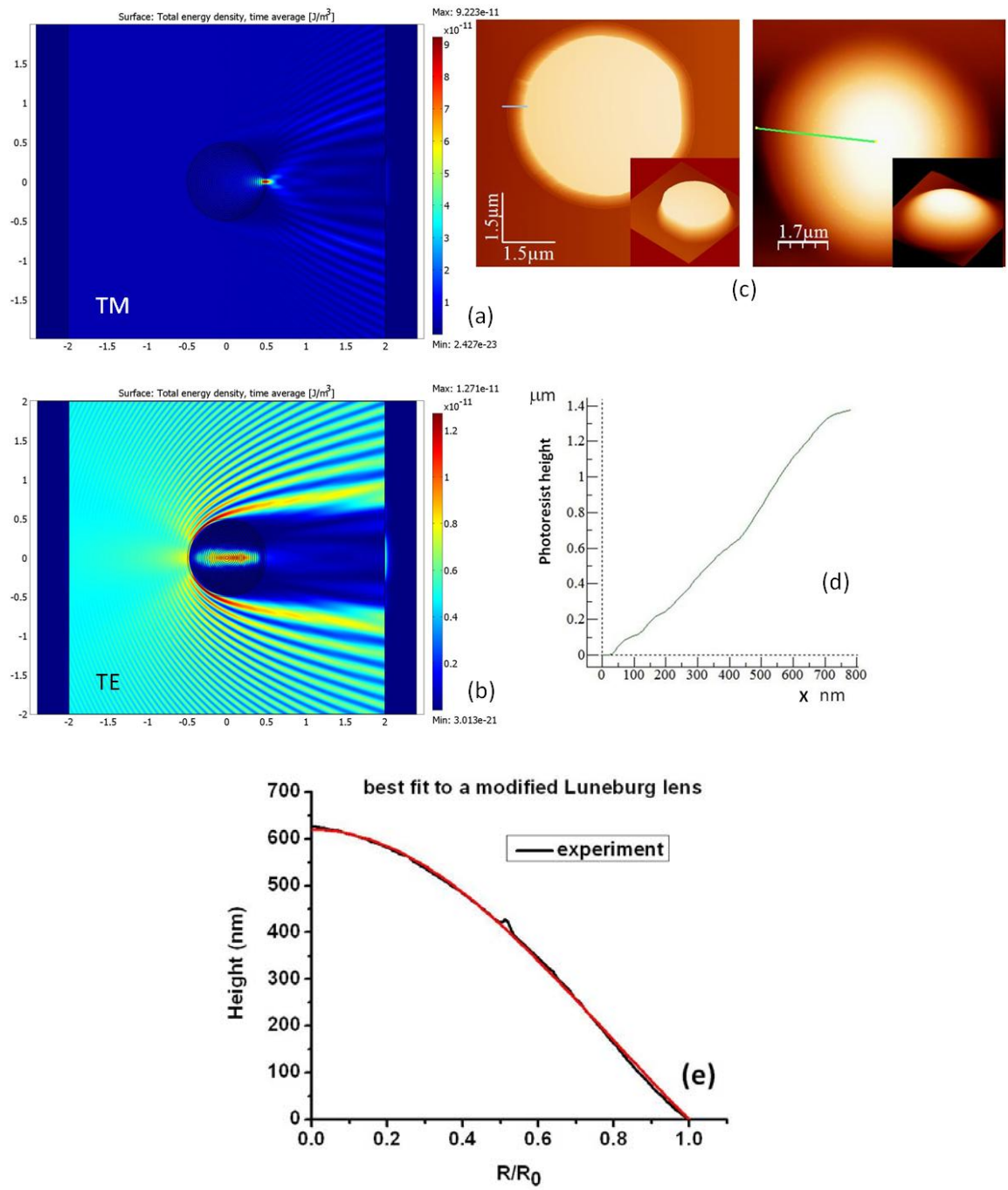

Fig. 2

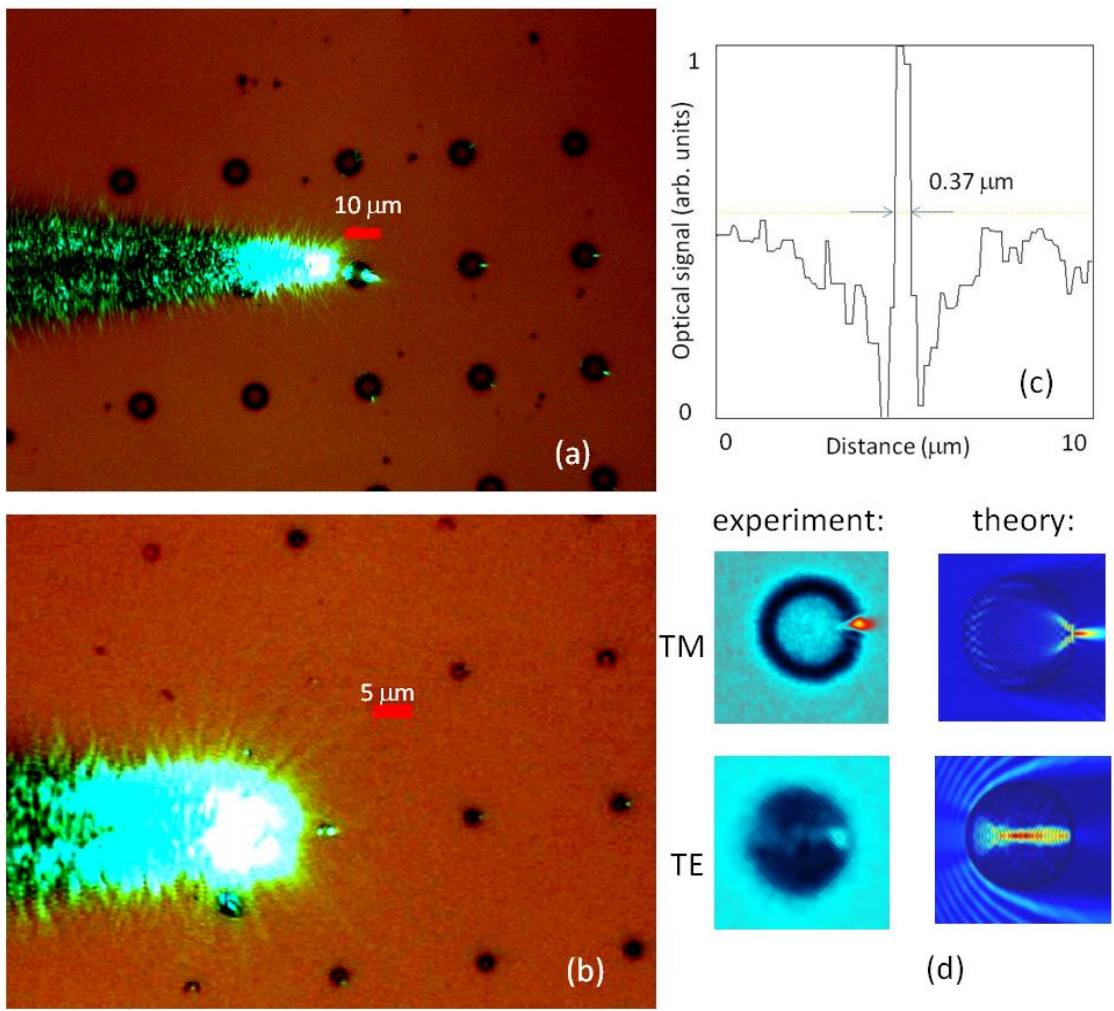

Fig. 3

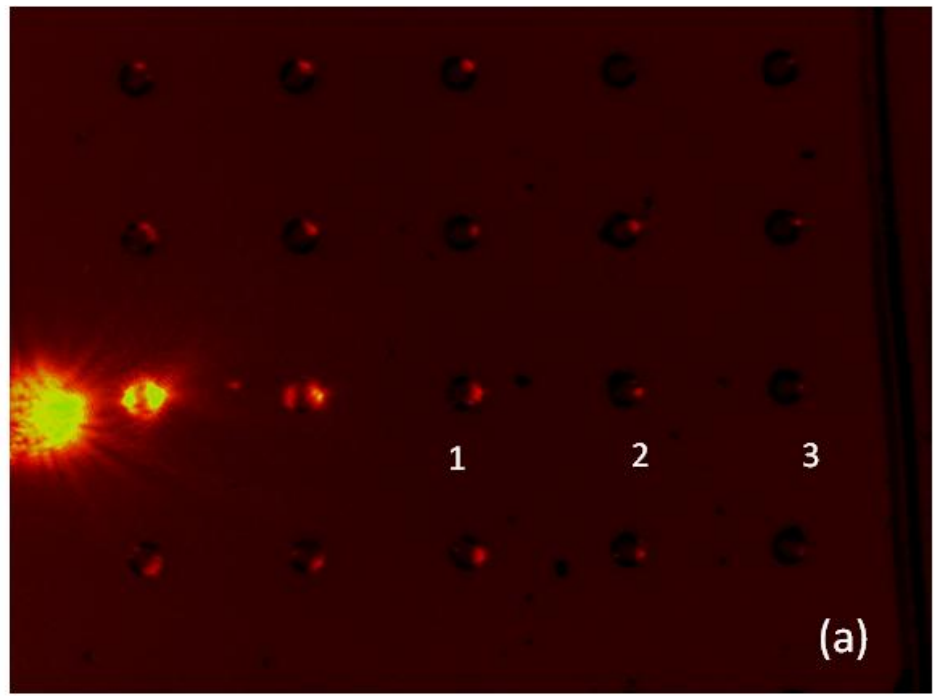

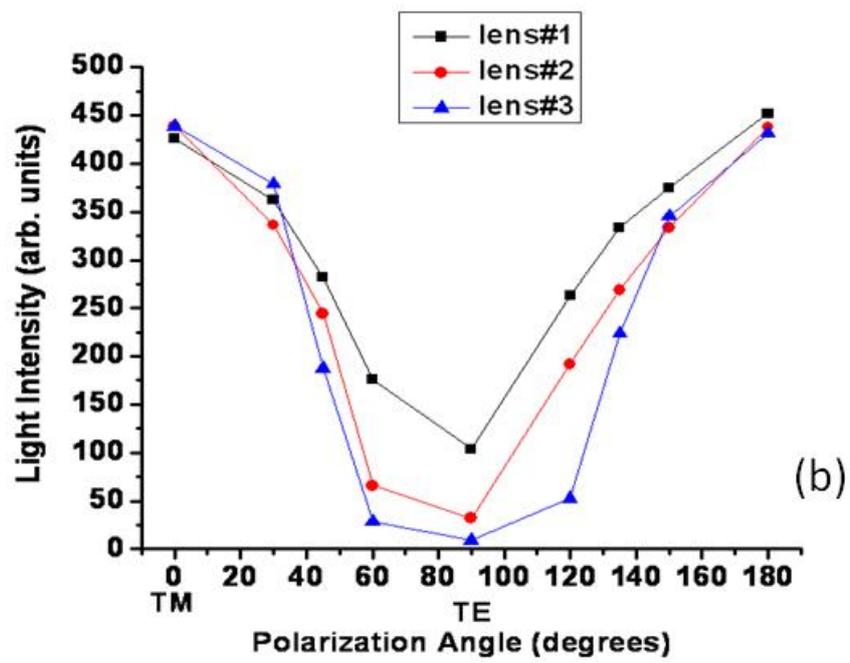

Fig. 4

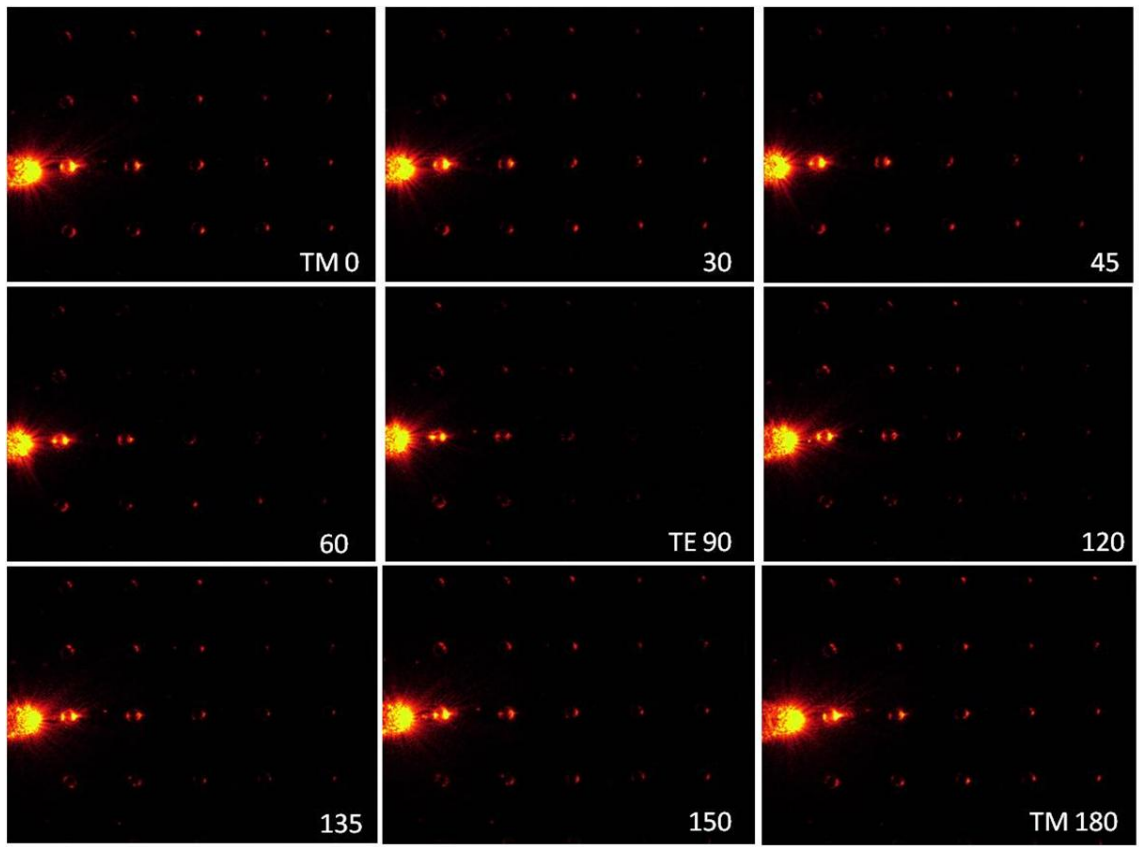

(c)

Fig.4

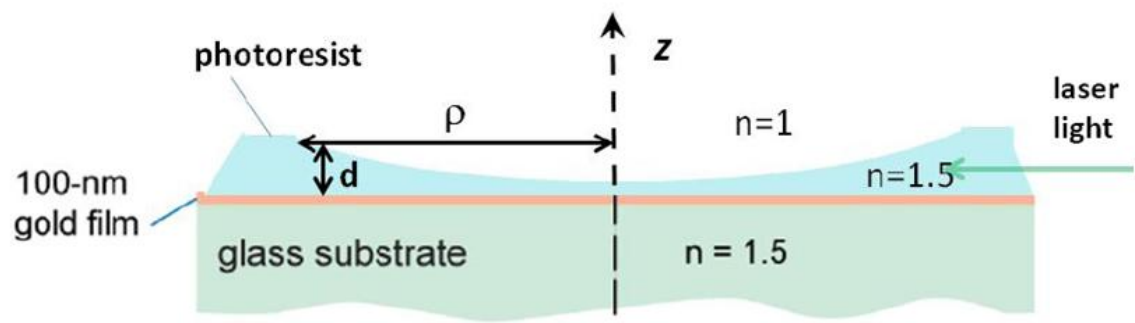

photoresist

$z$

$\rho$

n=1

laser
light

100-nm
gold film

d

n=1.5

glass substrate

n = 1.5

(a)

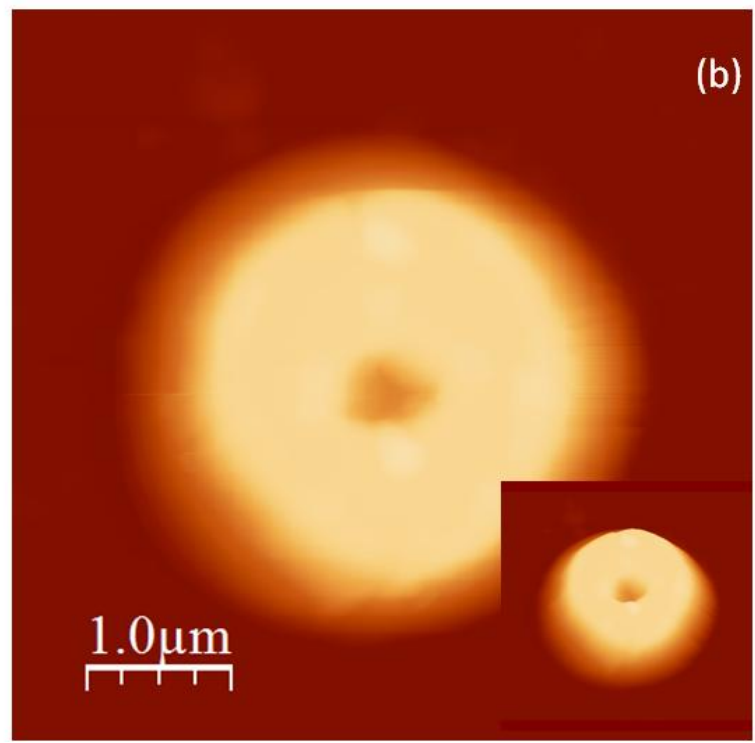

(b)

1.0μm

Fig. 5

　

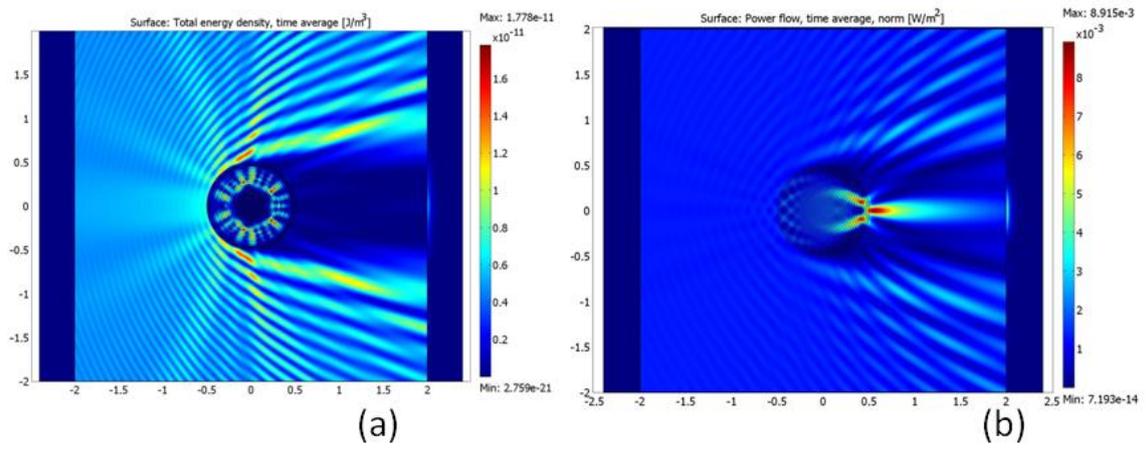

(a) (b)

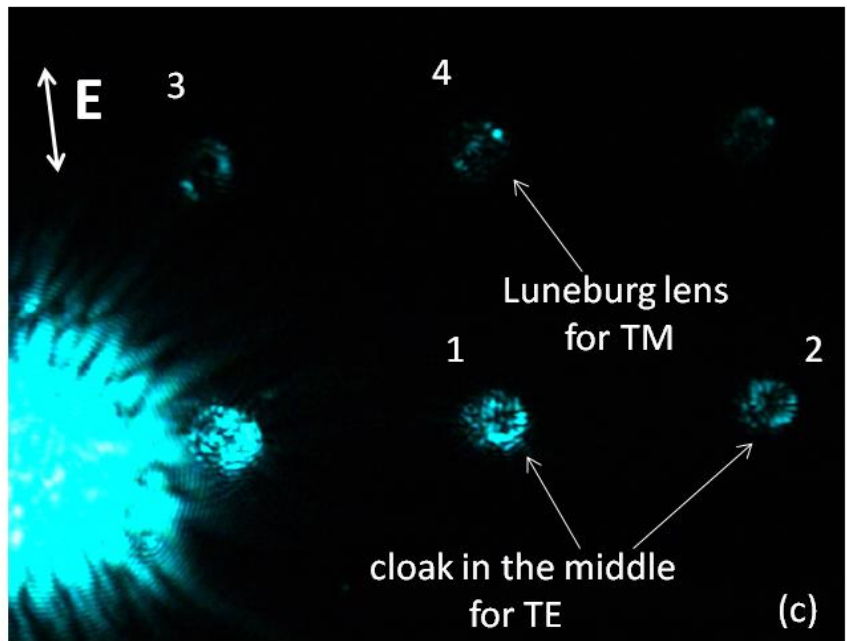

(c)

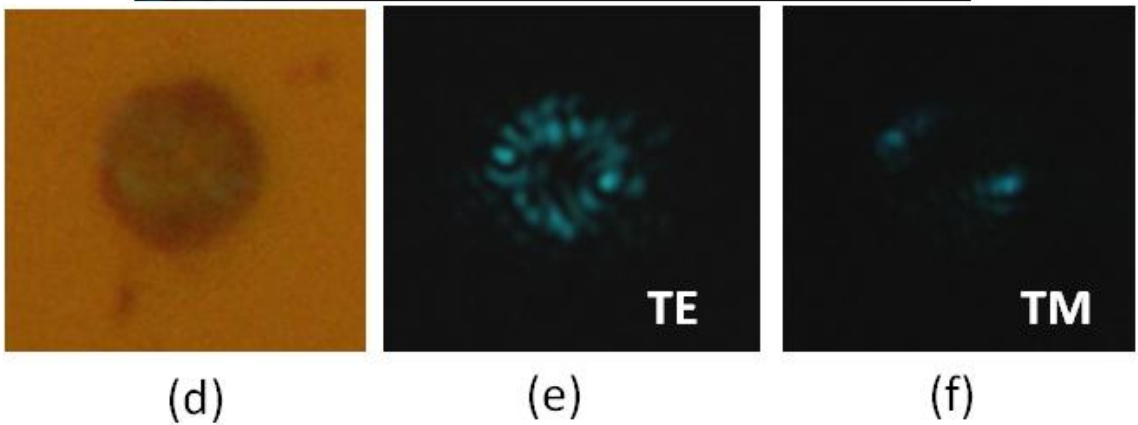

(d) (e) (f)

Fig. 6